\documentclass[aps,pra,amsmath,amsfonts,amssymb,twocolumn,showpacs]{revtex4}

\usepackage{amssymb}
\usepackage{graphicx,psfrag,color}
\usepackage{dcolumn}
\usepackage{bm}
\usepackage{mathbbol}

\begin{document}
\flushbottom

\title{Probabilistic quantum teleportation in the presence of noise}

\author{Raphael Fortes}
\author{Gustavo Rigolin}
\email{rigolin@ufscar.br}
\affiliation{Departamento de F\'isica, Universidade Federal de
S\~ao Carlos, 13565-905, S\~ao Carlos, SP, Brazil}

\date{\today}

\begin{abstract}
We extend the research program initiated in [Phys. Rev. A \textbf{92}, 012338 (2015)],
where we restricted our attention to noisy deterministic teleportation protocols, to noisy probabilistic (conditional)
protocols. Our main goal now is to study how we can increase the fidelity of the teleported state 
in the presence of noise by working with probabilistic protocols. We work with several scenarios involving 
the most common types of noise in realistic implementations of quantum communication tasks and find 
many cases where adding more noise to the probabilistic protocol increases considerably the fidelity
of the teleported state, without decreasing the probability of a successful run of the protocol. Also, there 
are cases where the entanglement of the channel connecting Alice and Bob leading to the greatest fidelity is not
maximal. Moreover, there exist cases where the optimal fidelity for the probabilistic protocols are greater than
the maximal fidelity ($2/3$) achievable by using only classical resources, while the optimal ones for the deterministic protocols 
under the same conditions lie below this limit. This result clearly
illustrates that in some cases we can only get a truly quantum teleportation if we use probabilistic instead 
of deterministic protocols.  
\end{abstract}

\pacs{03.65.Ud, 03.67.Bg, 05.40.Fb}

\maketitle

\section{Introduction} 

Quantum teleportation is a quantum communication task 
devised to transfer the quantum state of a physical system located at one place, 
say Alice's, to a different quantum system located at Bob's \cite{ben93,vai94,bra98,bow97,bos98,fur98}. 
Two important aspects of the teleportation protocol are related to the fact that it works 
without the knowledge of the quantum state to be teleported and that the physical system originally
described by this quantum state is not sent from Alice to Bob.
For the perfect functioning of the teleportation protocol Alice and
Bob need to share a maximally entangled state (maximally entangled quantum channel).
In this case the protocol works deterministically and with unity fidelity, i.e., every run of the protocol ends up with
Bob's system being described exactly by the
original state teleported by Alice. 
 
The requirement of a maximally entangled quantum channel connecting Alice 
and Bob is very difficult to achieve or maintain in practice since the inevitable 
presence of noise reduces the entanglement of the quantum state shared between them. 
In practical implementations of the teleportation protocol one can
either adopt entanglement distillation techniques \cite{ben96} or 
modify the original protocol in order to cope with the reduced level of entanglement 
\cite{guo00,agr02,rig06,bow01,alb02,lee02,tak12,hor00,ban02,yeo08,ban12,kno14,for15}.
In the first case Alice and Bob need to share several copies of partially entangled states 
before implementing an entanglement distillation protocol, whereby they obtain a
maximally entangled state at the expenses of many copies of partially entangled ones.
With this maximally entangled state, Alice and Bob are able to execute with success the original
teleportation protocol. In the second case, the partially entangled state is used as is and
the protocol is modified in order to achieve the greatest fidelity possible. In this last
case, we can divide all strategies in two groups. In the first group we have the deterministic
protocols and in the second group the probabilistic ones. 

The deterministic protocols \cite{bow01,alb02,lee02,tak12,hor00,ban02,yeo08,ban12,kno14,for15}
do not postselect any measurement outcome at Alice's and therefore are always ``successful'' in the
sense that any run of the protocol yields an output state to Bob, even if his state is not exactly
described by the original (input) state with Alice. The probabilistic protocols, on the other
hand, are not always successful as defined above since only certain measurement outcomes obtained by Alice are 
accepted. In the probabilistic protocols only those measurement outcomes leading to output states 
closest to the input are considered valid. In this way, by decreasing the success rate of the protocol
one increases the fidelity of the state with Bob (output) with respect to the input state \cite{guo00,agr02,rig06}.
It is worth mentioning that the probabilistic protocols given in Refs. \cite{guo00,agr02,rig06} assume the
non-maximally entangled state shared between Alice and Bob to be pure. 
 
In this article we want to extensively study 
probabilistic teleportation protocols where the quantum channel connecting Alice and
Bob are given by partially entangled mixed states. Our benchmarks are the optimal 
deterministic protocols given in Ref. \cite{for15}, i.e., we want to find situations
in which the reduction of the success rate (postselection) of the protocols in Ref.
\cite{for15} gives considerable improvements in the fidelity of the teleported state.

Being more specific, here we deal with several scenarios involving 
the four most common types of noise one faces when implementing a quantum communication protocol: 
the bit flip, the phase flip or phase damping, the depolarizing, and the amplitude damping noises. 
We also study situations in which the state to be teleported is also subjected to noise. We show that 
several of the interesting results obtained in \cite{for15} for the deterministic protocols are also present in the 
probabilistic case. For example, we show scenarios where more noise increases the fidelity
of the teleported state and where the entanglement of the quantum channel connecting Alice and Bob 
giving the greatest fidelity is not maximal. In addition to this, we show that 
there exist situations in which the probabilistic protocol outperforms the deterministic one in a very
important aspect. Indeed, we show that there are scenarios where the optimal fidelity for the probabilistic protocol is not only 
greater than the optimal one for the deterministic protocol, but the only one surpassing the maximal value 
achievable by using only classical resources. This fact is a clear indication that in some 
scenarios a truly quantum teleportation can only be obtained by using probabilistic protocols.

\section{Teleportation in the density matrix formalism}
\label{secI}

The mathematical concept needed to deal with noise and mixed states
is the density matrix and thus the first thing we need to do is to 
recast the original teleportation protocol using density matrices. 
This was done in full detail in Ref. \cite{for15} and here we only
give the key results necessary for the development of the ideas and 
concepts related to the probabilistic teleportation protocol. 

The input qubit's density matrix, i.e., Alice's qubit to be teleported to Bob,  
$|\psi\rangle_{in}= a|0\rangle + b|1\rangle$, with $|a|^2+|b|^2=1$, is  
\begin{equation}
\rho_{in} = |\psi\rangle_{in}\;_{in}\langle\psi | =
\left(
\begin{array}{cc}
|a|^2 & ab^* \\
a^*b & |b|^2
\end{array}
\right),
\end{equation}
where the subscript $in$ denotes ``input'' and $*$ complex conjugation.  
The initially noiseless entangled state shared between Alice and Bob, 
$|B_1^\theta\rangle=\cos\theta|00\rangle+\sin\theta|11\rangle$, has the following density matrix 
in the base $\{|00\rangle,|01\rangle,|10\rangle,|11\rangle\}$, 
\begin{equation}
\rho_{ch} = |B_1^\theta\rangle\langle B_1^\theta | =
\left(
\begin{array}{cccc}
\cos^2\theta & 0 & 0 & \sin\theta\cos\theta \\
0& 0 & 0 & 0 \\
0& 0 & 0 & 0 \\
\sin\theta\cos\theta & 0 & 0 & \sin^2\theta
\end{array}
\right).
\label{channel}
\end{equation}
Here $ch$ means ``channel'' and the first and second qubits are with Alice and Bob, respectively.
Note that $\theta$ is a free parameter that we can adjust to optimize the efficiency of the probabilistic teleportation.
When $\theta = \pi/4$ we have the maximally entangled state $|\Phi^+\rangle$, one of the four Bell states.
For any other value of $\theta\in [0,\pi/2]$ the entanglement of the state is not maximal, being zero for
$\theta = 0$ and $\pi/2$ \cite{woo98}.  

Using the above notation the global state
describing Alice's and Bob's qubits before the beginning of the protocol or the
action of noise is 
\begin{equation}
\rho = \rho_{in} \otimes \rho_{ch}.
\label{step1}
\end{equation}

The protocol begins by Alice making a projective measurement on her
two qubits (the input state and her share of the entangled state). These qubits
are projected onto one the four states listed below that form a complete basis,
\begin{eqnarray}
|B_1^\varphi\rangle&=&\cos\varphi|00\rangle+\sin\varphi|11\rangle, \label{B1} \\
|B_2^\varphi\rangle&=&\sin\varphi|00\rangle-\cos\varphi|11\rangle, \\
|B_3^\varphi\rangle&=&\cos\varphi|01\rangle+\sin\varphi|10\rangle, \label{B3} \\
|B_4^\varphi\rangle&=&\sin\varphi|01\rangle-\cos\varphi|10\rangle. \label{B4}
\end{eqnarray}
In the original protocol $\varphi=\pi/4$, with those states becoming the usual Bell states,
$|\Phi^+\rangle, |\Phi^-\rangle, |\Psi^+\rangle$, and $|\Psi^-\rangle$. Here $\varphi$
is also a free parameter that is chosen to maximize the efficiency of the probabilistic 
teleportation. 
The projectors associated with these four states are,
\begin{equation}
P_j^\varphi=|B_j^\varphi\rangle\langle B_j^\varphi|, \hspace{.5cm} j=1,2,3,4.
\end{equation}
After this measurement the 
global state, Eq. (\ref{step1}), changes to
\begin{equation}
\tilde{\rho}_j = \frac{P_j^\varphi \rho P_j^\varphi}{\mbox{Tr}[{P_j^\varphi \rho}]} 
\label{nine}
\end{equation}
with probability
\begin{equation}
Q_j(|\psi\rangle_{in}) = \mbox{Tr}[{P_j^\varphi \rho}],
\label{prob}
\end{equation}
where $\mbox{Tr}$ is the trace operation. Note that we have 
explicitly written the dependence of $Q_j$ on the input state
$|\psi\rangle_{in}$. Only for maximally entangled channels this 
probability is independent of the initial state \cite{guo00,agr02,rig06}. 

In the second step of the protocol Alice tells Bob, using a classical communication channel,
which $|B_j^\varphi\rangle$ she measured. After receiving this information,
Bob knows that his state is now described by
\begin{equation}
\tilde{\rho}_{_{B_j}} = \mbox{Tr}_{12}[\tilde{\rho}_j] = \frac{\mbox{Tr}_{12}[P_j^\varphi \rho P_j^\varphi]}{Q_j(|\psi\rangle_{in})}, 
\end{equation}
where $\mbox{Tr}_{12}$ denotes the partial trace on qubits $1$ and $2$ (those with Alice). 

In the third and last step of the protocol Bob implements a unitary operation $U_j$ on his state
in order to recover exactly the teleported state.
After this unitary operation the final state with Bob is given by 
\begin{equation}
\rho_{_{B_j}}= U_j\tilde{\rho}_{_{B_j}}U_j^\dagger =  \frac{U_j\mbox{Tr}_{12}[P_j^\varphi \rho P_j^\varphi]U_j^\dagger}{Q_j(|\psi\rangle_{in})}.
\label{twelve}
\end{equation}
It is worth noting that the unitary operation that Bob implements depends on Alice's measurement result 
and on the quantum channel used in the protocol. For $\rho_{ch}$ given by Eq.~(\ref{channel}),
$U_1=\mathbb{1}$, $U_2=\sigma_z, U_3=\sigma_x$, and $U_4=\sigma_z\sigma_x$, where
$\mathbb{1}$ is the identity matrix and $\sigma_z$ and $\sigma_x$ the standard Pauli matrices.

\section{Teleportation in the presence of noise}
\label{noise}

The operator-sum representation formalism \cite{kra83,nie00} is the mathematical concept we need to
model in the simplest way the action of noise on a qubit. 
The key concept behind this formalism is that 
the noise can be described only by quantum operations belonging to the qubit's Hilbert space. 
The operators $E_k$ representing a particular kind of noise are called Kraus operators 
and for trace preserving operations (conservation of probability) 
they must obey the condition
\begin{equation}
\sum_{j=1}^nE_j^\dagger E_j = \mathbb{1},
\end{equation}
where $\mathbb{1}$ is the identity operator 
acting on the qubit's Hilbert space and $1\leq n\leq 4$. 
The action of the noise on the 
qubit $k$, described by the density matrix $\rho_k$, is
\begin{equation}
\rho_k \rightarrow \varrho_k = \sum_{j=1}^nE_j \rho_k E_j^\dagger. 
\label{actionOfnoise}
\end{equation}

Throughout this section we follow closely the notation and presentation of Ref. \cite{for15}
and just list the most common types of noise
we usually find in any realistic modeling of a qubit lying in a 
noisy environment. We consider four types of noise, namely, the bit flip,
the phase flip or phase damping, the depolarizing, and the amplitude
damping channels. The physical meaning of each one of these noise channels
are extensively discussed in Ref. \cite{kra83,nie00} and a brief discussion
can be found in Ref. \cite{for15}. 
The Kraus operators representing the action of those noise channels are
given in Tab. \ref{table1}.

\begin{table}[!ht]
\caption{\label{table1} 
Here $p \in [0,1]$ is the probability that the noise has acted on the qubit and
$\sigma_j, j=x,y,$ and $z$, are the standard Pauli matrices.}
\begin{ruledtabular}
\begin{tabular}{ll}
 & \\
Bit flip & $E_1 = \sqrt{1-p}\;\mathbb{1}, \hspace{.5cm} E_2 = \sqrt{p}\;\sigma_x.$ \\
 & \\
\hline
 & \\
Phase flip & $E_1 = \sqrt{1-p}\;\mathbb{1}, \hspace{.5cm} E_2 = \sqrt{p}\;\sigma_z.$ \\
(Phase damping) &  \\
\hline
 & \\
Depolarizing & $E_1 = \sqrt{1-3p/4}\;\mathbb{1}, \hspace{.2cm} E_2 = \sqrt{p/4}\;\sigma_x,$ \\
             & $E_3 = \sqrt{p/4}\;\sigma_y,  \hspace{.7cm} E_4 = \sqrt{p/4}\;\sigma_z.$ \\
 & \\
\hline
 & \\
Amplitude damping & 
$
E_1 =
\left(
\begin{array}{cc}
1 & 0 \\
0 & \sqrt{1-p}
\end{array}
\right), 
\hspace{.1cm} 
E_2 = 
\left(
\begin{array}{cc}
0 & \sqrt{p} \\
0 & 0
\end{array}
\right).
$ \\
  & 
\end{tabular}
\end{ruledtabular}
\end{table}

Assuming that each qubit in the teleportation protocol is acted on by 
noise in an independent way, the global density matrix describing the initial state, Eq.~(\ref{step1}), 
becomes \cite{for15}
\begin{equation}
\varrho =  
\sum_{i=1}^{n_{\!_I}}\sum_{j=1}^{n_{\!_A}}\sum_{k=1}^{n_{\!_B}}
E_{ijk}(p_{\!_I},p_{\!_A},p_{\!_B})\rho E_{ijk}^\dagger(p_{\!_I},p_{\!_A},p_{\!_B}).
\label{noiseRho}
\end{equation}
Equation (\ref{noiseRho}) is obtained by 
applying Eq.~(\ref{actionOfnoise}) to each one of the qubits in Eq.~(\ref{step1}). 
Here $E_{ijk}(p_{\!_I},p_{\!_A},p_{\!_B})=E_i(p_{\!_I})\otimes F_j(p_{\!_A})\otimes G_k(p_{\!_B})$,
where $E_i(p_{\!_I})=E_i(p_{\!_I})\otimes\mathbb{1}\otimes\mathbb{1}, 
F_j(p_{\!_A})=\mathbb{1}\otimes F_j(p_{\!_A})\otimes\mathbb{1}$,
and $G_k(p_{\!_B})=\mathbb{1}\otimes\mathbb{1} \otimes G_k(p_{\!_B})$ are, respectively, 
the Kraus operators related to the noise acting on the input qubit and Alice's and Bob's qubits 
of the quantum channel. In order to keep track that in general 
different types of noises can act during different times (probabilities) we explicitly
show the dependence of the Kraus operators on those probabilities: $p_{\!_I}, p_{\!_A}$, and $p_{\!_B}$. 
The density matrix $\varrho$, Eq.~(\ref{noiseRho}), should be used instead of $\rho$  
in Eqs.~(\ref{nine}) to (\ref{twelve}) to get the relevant
quantities needed to analyze the probabilistic teleportation protocol
in the presence of noise.

\section{Rate of success and efficiency of the noisy probabilistic teleportation}
\label{fidelity}

In the presence of noise \cite{for15}, or when we deal with non-maximally entangled channels \cite{guo00,agr02,rig06},
the probability $Q_j$ of Alice measuring a determined generalized Bell state $|B_j^\varphi\rangle$
depends on the input state $|\psi\rangle_{in}$ to be teleported. Thus, in order to be as general as possible
and to get results that are independent of a specific input state, we assume a uniform probability 
distribution 
\begin{equation}
P_X(x)=\mathcal{P}(|\psi\rangle_{in})
\end{equation}
for those input states \cite{for15}.  Here $X$ is a continuous random variable whose possible values $x$ are all pure qubits that
define the sample space $\Omega$. We will work with  
a probability distribution $P_X(x)$ that is normalized,  
\begin{equation}
\int_{\Omega} P_X(x) dx=\int_{\Omega}\mathcal{P}(|\psi\rangle_{in})d|\psi\rangle_{in} = 1,
\label{normalization1}
\end{equation}
and, as we said, uniform (Haar measure), i.e., $P_X(x)$ is the same (constant) for all $x$. With this choice for
$P_X(x)$ all qubits have equal chances of being picked by Alice at each run of the protocol.

Being more specific, writing an arbitrary qubit as
\begin{equation}
|\psi\rangle = \alpha|0\rangle + \beta e^{i\gamma}|1\rangle,
\label{relative}
\end{equation}
with $\alpha$, $\beta$, and $\gamma$ positive real numbers such that $\alpha^2+\beta^2=1$ and $\gamma\in [0,2\pi]$,
we can choose $\alpha^2$ and $\gamma$ as our independent variables. 
With this notation $\mathcal{P}(|\psi\rangle_{in})=\mathcal{P}(\alpha^2,\gamma)$ and 
the normalization condition, Eq.~(\ref{normalization1}), becomes
\begin{equation}
\int_{0}^{2\pi}\int_{0}^{1}\mathcal{P}(\alpha^2,\gamma)\mathrm{d}\alpha^2\mathrm{d}\gamma =1.
\label{normP}
\end{equation}
For a uniform probability distribution ($\mathcal{P}(\alpha^2,\gamma)$ constant) Eq.~(\ref{normP}) implies 
\begin{equation}
\mathcal{P}(\alpha^2,\gamma)=\frac{1}{2\pi}.
\label{uniform}
\end{equation}

We also have a discrete variable $J$ whose values can be $j=1,2,3,$ and $4$ (or $j=\Phi^+,\Phi^-,\Psi^+,$ and $\Psi^-$), with each $j$ representing one
of the four possible generalized Bell states $|B_j^\varphi\rangle$. The probability to measure a given $|B_j^\varphi\rangle$
is written as $P_J(j)$. 
The conditional probability $P_{J|X}(j|x)$ is the chance of Alice measuring the Bell state $j$ if
she teleports the input state $x$ and it is given by Eq. (\ref{prob}),
\begin{equation}
P_{J|X}(j|x) = 
Q_j(|\psi\rangle_{in}).
\end{equation}

To determine $P_J(j)$ we first determine the joint probability distribution
$P_{XJ}(x,j)$ by applying the well-known result of probability theory that 
says that 
\begin{equation}
P_{XJ}(x,j)=P_{JX}(j,x)=P_{X}(x)P_{J|X}(j|x). 
\label{relationP}
\end{equation}
Thus, using Eq.~(\ref{relationP}) we get  
\begin{equation}
P_{XJ}(x,j) = \mathcal{P}(|\psi\rangle_{in})Q_j(|\psi\rangle_{in}).
\end{equation}
Now, since the marginal probability distribution is 
$P_J(j)=\int_\Omega P_{XJ}(x,j) dx$ we have
\begin{equation}
P_{J}(j) = \int_\Omega \mathcal{P}(|\psi\rangle_{in})Q_j(|\psi\rangle_{in}) d|\psi\rangle_{in}.
\label{marginalJ}
\end{equation}

At last, using Eq.~(\ref{marginalJ}) and again Eq.~(\ref{relationP}) with the roles of $X$
and $J$ interchanged we get
\begin{eqnarray}
P_{X|J}(x|j) &=& 
\frac{P_{XJ}(x,j)}{P_{J}(j)} 
\nonumber \\
&=&  
\frac{\mathcal{P}(|\psi\rangle_{in})Q_j(|\psi\rangle_{in})}
{\int_\Omega \mathcal{P}(|\psi\rangle_{in})Q_j(|\psi\rangle_{in}) d|\psi\rangle_{in}}.
\label{finalP}
\end{eqnarray}
Equations (\ref{marginalJ}) and (\ref{finalP}) are the relevant probability distributions we need to 
quantitatively analyze the probabilistic teleportation protocol. Indeed, 
$P_{J}(j)$ is the probability to measure a given generalized Bell state $j$ given a certain distribution for the input states
and it can be interpreted as the average chance of measuring $|B_j^\varphi\rangle$,
\begin{equation}
\overline{Q}^j= P_{J}(j) = \int_\Omega \mathcal{P}(|\psi\rangle_{in})Q_j(|\psi\rangle_{in}) d|\psi\rangle_{in}.
\label{averageQ}
\end{equation}
This quantity is independent of $|\psi\rangle_{in}$ and it is referred to here as the success rate or probability
of success of the probabilistic teleportation protocol when we postselect a particular measurement result $j$. 
$P_{X|J}(x|j)$, as we will show shortly, is the quantity we need to compute the input-state-independent efficiency of the protocol once we fix
our attention to a given measurement outcome $j$. $P_{X|J}(x|j)$ is the probability distribution of the input 
states $x$ when we consider only (postselect) those measurement results at Alice's yielding the 
same generalized Bell state $j$.

To quantify the efficiency of the probabilistic teleportation protocol we use the fidelity \cite{uhl76}.
Since in our analysis the input state (our benchmark) is initially pure, the fidelity is
\begin{equation}
F_j = \mbox{Tr}[\rho_{in}\varrho_{_{B_j}}]=\,_{in}\langle \psi | \varrho_{_{B_j}} | \psi \rangle_{in},
\label{Fj}
\end{equation}
where $\varrho_{_{B_j}}$ is the state with Bob at the end of a run of the protocol, Eq.~(\ref{twelve}), with $\rho$ 
changed to the noisy state $\varrho$, Eq.~(\ref{noiseRho}).
Equation (\ref{Fj}) ranges from zero to one, being one whenever the output state ($\varrho_{_{B_j}}$) is equal 
(up to an irrelevant global phase) to the input ($| \psi \rangle_{in}$) and zero whenever the two states are orthogonal.

Since $F_j$ depends on the input state $|\psi\rangle_{in}$ we must average $F_j$ over all possible input states 
to obtain a quantitative description of the efficiency of the protocol that is independent 
of $|\psi\rangle_{in}$. Since the probability distribution for $|\psi\rangle_{in}$ within a given fixed choice
of measurement result $j$ is $P_{X|J}(x|j)$, Eq.~(\ref{finalP}), we get
\begin{eqnarray}
\overline{F}^j &=& \int_{\Omega}F_j(x)P_{X|J}(x|j)dx \nonumber \\
&=&
\frac{\int_\Omega F_j(| \psi \rangle_{in})\mathcal{P}(|\psi\rangle_{in})Q_j(|\psi\rangle_{in})d|\psi\rangle_{in}}
{\int_\Omega \mathcal{P}(|\psi\rangle_{in})Q_j(|\psi\rangle_{in}) d|\psi\rangle_{in}}
\label{FinalFj} 
\end{eqnarray}
for the efficiency of the probabilistic teleportation protocol when postselecting the 
measurement result $j$. Note that if we consider all measurement outcomes as acceptable we 
recover the deterministic protocol of Ref. \cite{for15}. In the present notation the quantity
employed in Ref. \cite{for15} to quantify the efficiency of the deterministic protocol reads
\begin{equation}
\langle \overline{F} \rangle = \sum_{j=1}^{4}P_J(j)\overline{F}^j = 
\int_\Omega \overline{F}(|\psi\rangle_{in}) \mathcal{P}(|\psi\rangle_{in})d|\psi\rangle_{in},
\label{deterministicF}
\end{equation}
where $\overline{F}(|\psi\rangle_{in})= \sum_j^4Q_j(|\psi\rangle_{in})F_j(| \psi \rangle_{in})$. 
One of our goals in this work is to optimize Eq.~(\ref{FinalFj}) such that 
$\overline{F}^j > \langle\overline{F}\rangle$, with $\langle\overline{F}\rangle$ being the optimal
efficiency of the deterministic protocol. In this case 
the probabilistic protocol outperforms the deterministic one in terms of efficiency, i.e.,
the teleported state with Bob is closer to the original one with Alice. The price we pay is a reduction of 
the probability of success since we have to discard measurement results different from $j$.

Summing up, Eqs.~(\ref{marginalJ}) and (\ref{FinalFj}) are the relevant expressions employed here to quantify,
respectively, the probability of success and the efficiency (fidelity) of the probabilistic teleportation protocol;
and Eq.~(\ref{deterministicF}), the efficiency of the deterministic protocol, is the benchmark we want to surpass by optimizing
(\ref{FinalFj}). With these equations and the ideas and concepts here developed, 
we are now ready to move on to the quantitative analysis of the interplay between 
probability of success and efficiency for several noise scenarios in the next section.

\section{Results}
\label{results}

\begin{figure}[!ht]
\includegraphics[width=8.5cm]{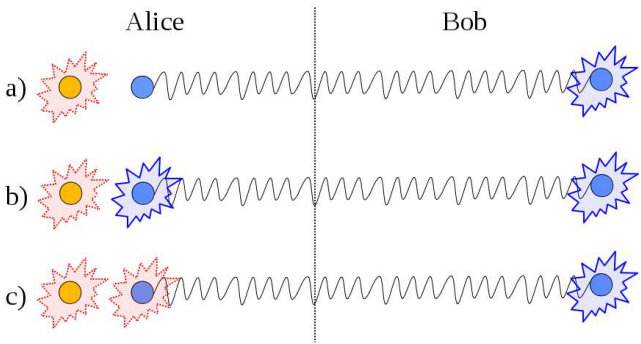}
\caption{\label{fig0}(color online) The three noise scenarios studied here.
a) Noise acting on Alice's input qubit and on Bob's output qubit. b) Noise 
acting on the input qubit and the same type of noise acting on the qubits of the channel.
c) Noise on Bob's qubit and the same type of noise on Alice's qubits. See text for details.}
\end{figure}

We study the efficiency of the probabilistic teleportation protocol in the three noise scenarios 
presented in Ref. \cite{for15} for the deterministic protocol. The first one 
assumes that only Bob's qubit is subjected to noise in addition to 
the input qubit, which can suffer the action of the same or of a different type of noise (see Fig.~\ref{fig0}-a). 
Note that by choosing Alice's qubit of the quantum channel to be acted on by noise 
instead of Bob's leads to the same results \cite{for15}. 
The second scenario we investigate is the one in which the entangled state shared by Alice and Bob 
are subjected to the same kind of noise during the same time, 
while the input qubit can suffer the action of any one of the four types
of noises explained in Sec. \ref{noise} (see Fig. \ref{fig0}-b). This situation occurs when 
the quantum channel is created by a third
party symmetrically located between Alice and Bob such that both qubits of the channel
find similar noisy environments during their flights to Alice and to Bob. In the notation of 
Sec. \ref{noise} this implies that $p_{\!_A}=p_{\!_B}=p$. 
The third scenario we investigate is the one in which all Alice's qubits 
are subjected to the same kind of noise while Bob's qubit can suffer the action of 
the same or of a different noise (see Fig. \ref{fig0}-c). 
This scenario is relevant when it is 
Alice that generates the entangled channel. In such a case the input qubit 
and her share of the entangled state lie
in the same environment and therefore are acted on by the same noise and during the same time.
In the notation of Sec. \ref{noise} it means that $p_{\!_I}=p_{\!_A}=p$.

Before we continue it is important to review and adapt the notation introduced in Ref. \cite{for15} 
designed to concisely label which qubits are subjected to a particular kind of noise in the expressions for
the probability of success and efficiency that will follow. In this notation any quantity that depends
on the arrangement of the types of noise acting on the three qubits of the teleportation protocol is written
with three subindexes representing each type of noise. For example, the average probability (probability of success)
of Alice obtaining a given Bell state $j$ for a given noise configuration is written as 
$\overline{Q}^j_{_{X,\varnothing,Y}}$, where the first subindex denotes that the input qubit
is subjected to noise $X$, the second one represents that Alice's qubit of the quantum channel lies in a noiseless
environment, and the third subindex tells us that Bob's qubit is subjected to noise $Y$.

\subsection{Scenario $1$}

\begin{figure}[!ht]
\includegraphics[width=8.5cm]{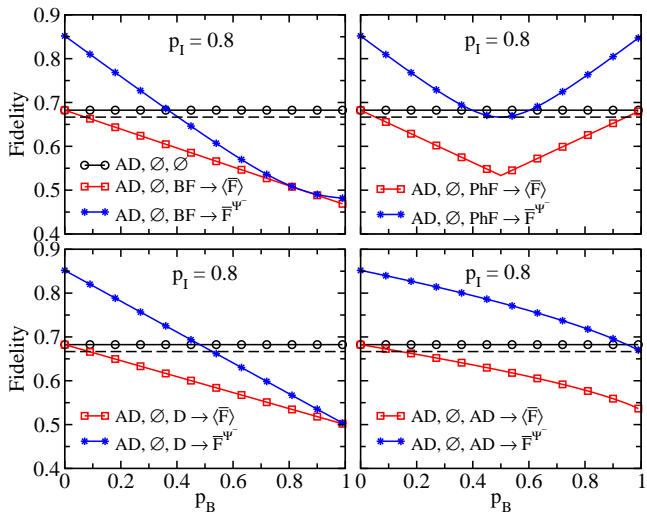}
\caption{\label{fig1}(color online) Optimal efficiencies (average fidelities) as a function of $p_{\!_B}$ for the deterministic protocol
when only the input qubit is subjected to noise (circle-black curves), for the deterministic protocol
when both the input and Alice's qubit are subjected to noise (square-red curves), and for the probabilistic protocol with both
input and Bob's qubit acted on by noise (star-blue curves). The several noise arrangements and the value of $p_{\!_I}$ are given in the figure.
The dashed-black curves mark the classical limit $2/3$ for the fidelity. The probabilistic optimal efficiencies are those obtained
by post\-selecting $|\Psi^-\rangle$. The optimal $\overline{F}^j$ are the same for all possible measurement
results if Bob's qubit is subjected to the bit flip, phase flip, or depolarizing noise. When Bob's qubit
is subjected to the amplitude damping noise, the best results are obtained only if Alice postselects 
$|\Psi^+\rangle$ or $|\Psi^-\rangle$. For the range of $p_{\!_B}$ in which $\overline{F}^j>2/3$, 
the rate of success $\overline{Q}^j$ for all the $4$ cases are of the order of
$10\%$, with lowest values being around $5\%$. Here and in the following figures all plotted quantities are dimensionless.}
\end{figure}

The first scenario we investigate is the one depicted in Fig. \ref{fig0}-a, where only the input and Bob's 
qubit are subjected to noise. The input qubit can suffer the action of any one of the four types of noise 
given in Sec. \ref{noise} as well as Bob's. We thus have $16$ possible noise arrangements. For each one 
of these arrangements we have optimized all four $\overline{F}^j$, Eq.~(\ref{FinalFj}),
as a function of $\theta$ and $\varphi$,
variables related, respectively, to the initial  
entanglement (prior to the action of noise) of the quantum channel and to the projective measurement implemented by Alice. 
See Eqs.~(\ref{channel}) and (\ref{B1})-(\ref{B4}). Comparing within a given noise arrangement 
the four optimal $\overline{F}^j$ with the optimal $\langle \overline{F} \rangle$,
the efficiency for the deterministic protocol (Eq.~(\ref{deterministicF})),  
we noted that only $4$ out of these $16$ possibilities yielded at least one $j$ such that 
$\overline{F}^j > \langle \overline{F} \rangle$ (See Fig. \ref{fig1}). 

A common feature among these $4$ cases is the action of the amplitude damping noise on the input qubit.
Indeed, if the input qubit is subjected to any other type of noise, the optimal probabilistic efficiency satisfies 
$\overline{F}^j = \langle \overline{F} \rangle$ for all $j$. 
Another feature shared by these $4$ cases is the fact that 
the initial entanglement of the quantum channel connecting Alice and Bob 
giving the optimal efficiency is not maximal whenever $p_{\!_I}\neq 0$. 
This is a situation 
where \textit{less} entanglement leads to \textit{more} efficiency. This same feature is seen for the deterministic
protocol when we also deal with the amplitude damping noise \cite{ban12,for15}. For the other $12$ cases in this scenario,
the initial entanglement leading to the optimal efficiency is maximal for both the deterministic and probabilistic protocols.

In Fig. \ref{fig1} we show for these $4$ cases the optimal values
for the efficiency of the probabilistic protocol versus the optimal one for the deterministic protocol when 
$p_{\!_I}=0.8$ and for all values of $p_{\!_B}$. For other values of $p_{\!_I}$ we have the same qualitative behavior.
Looking at Fig. \ref{fig1} we notice another important feature
shared by all these $4$ cases. We can always find a critical value for $p_{\!_B}$ below which 
$\overline{F}^j_{AD,\varnothing,Y}>\langle \overline{F} \rangle_{AD,\varnothing,Y}$ and at the same time
$\overline{F}^j_{AD,\varnothing,Y}>\langle \overline{F} \rangle_{AD,\varnothing,\varnothing}$. In other words, 
the optimal probabilistic efficiency is not only greater than the optimal one for the deterministic protocol under the
same noise conditions but also greater than the optimal deterministic protocol efficiency when Bob's qubit is not subjected to noise.
This is an instance where \textit{more} noise leads to \textit{more} efficiency. 

We also have two interesting results in those $4$ noise arrangements. The first one occurs for high values of $p_{\!_I}$. Under
this condition the fidelities of the deterministic protocols almost always lie below $2/3$, being slightly above this value
only for very small values of $p_{\!_B}$ (See Fig. \ref{fig1}). 
Any fidelity below $2/3$ can be achieved using only classical resources (no need for entanglement) and the
teleportation protocol is considered genuinely quantum for a uniform probability distribution (Haar measure) of input states
only if we have fidelities greater than $2/3$ \cite{bra00}. 
On the other hand, for the probabilistic protocol we can significantly surpass the classical limit for a considerable
range of values for $p_{\!_B}$. For the noise arrangements where Bob's qubit is subjected to either the phase flip or
the amplitude damping noise, we obtain for almost all values of $p_{\!_B}$ fidelities greater than $2/3$, clearly illustrating
that the probabilistic protocols are the only ones leading to a truly quantum teleportation.
The second interesting result occurs when noise is unavoidable and Bob can choose in which noisy
environment to keep his qubit. In such a case subjecting his qubit to a different kind of noise than that 
acting on the input qubit can be beneficial. This does not change the probability of success, since it only depends
on what is happening at Alice's, but increases the efficiency of the protocol. For example, looking at
Fig. \ref{fig1} we see that $\overline{F}^j_{AD,\varnothing,PhF} > \overline{F}^j_{AD,\varnothing,AD}$ for
high values of $p_{\!_B}$. This is an illustration that \textit{different} noises lead to 
\textit{more} efficiency.

\subsection{Scenario $2$}

Let us now move to the case where both qubits of the quantum channel are acted on by the same noise while
Alice's input qubit is subjected to the same or a different type of noise (see Fig. \ref{fig1}-b). In this
scenario $p_{\!_A}=p_{\!_B}=p$ and similarly to the previous one we have $16$ possible combinations of noise. In order to optimize the 
efficiency of the probabilistic protocols we proceeded in the same
way as explained in scenario $1$. Out of these
$16$ arrangements, only those $7$ in which the amplitude damping noise is present 
yield probabilistic protocols with optimal efficiencies
greater than the optimal ones for the deterministic protocols.  
For these $7$ cases the initial entanglement of the entangled state shared by Alice and Bob 
giving the optimal efficiency is not maximal, similarly to what we have seen in scenario $1$. 
This is again a situation where \textit{less} entanglement leads to \textit{more} efficiency. 

\begin{figure}[!ht]
\includegraphics[width=8.5cm]{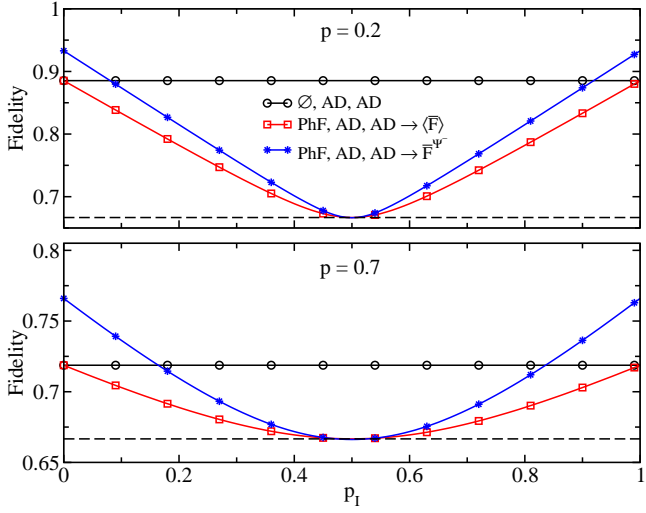}
\caption{\label{fig2}(color online)  Optimal efficiencies (average fidelities) as a function of $p_{\!_I}$ for the deterministic protocol
when only the quantum channel is subjected to noise (circle-black curves) and for the deterministic (square-red curves) and probabilistic (star-blue curves)
protocols when all three qubits are subjected to noise. The noise arrangements and the value of $p=p_{\!_A}=p_{\!_B}$ are in the figures.
The dashed-black curves mark the classical limit $2/3$ for the fidelity. The optimal efficiencies for the probabilistic protocol
are those obtained by post\-selecting $|\Psi^-\rangle$. Note that for almost all values of $p_{\!_I}$ the probabilistic protocol outperforms the
deterministic one. Moreover, for small ($\lesssim 0.1$) and high ($\gtrsim 0.9$) values of
$p_{\!_I}$ the probabilistic protocol gives a better result even when compared to the deterministic protocol 
in which no noise acts on the input qubit (circle-black curves).  The success rate $\overline{Q}^j$ for this to happen 
is of the order of $0.5\%$ for the two values of $p$ shown above. This is another example where \textit{more} 
noise leads to \textit{more} efficiency.}
\end{figure}

In Fig. \ref{fig2} we show the optimal fidelities for the deterministic and probabilistic protocols 
for the noise arrangement in which the quantum channel is subjected to the amplitude
damping noise and the input qubit to the phase flip noise ($PhF,AD,AD$). But to one feature the same qualitative behavior 
seen in this case are also present when the input qubit is subjected to the bit flip ($BF,AD,AD$) and depolarizing ($D,AD,AD$) noises. 
The only notable qualitative difference is related to the fact that while for the $PhF,AD,AD$ case the optimal efficiencies are symmetrical 
with respect to the line $p_{\!_I}=0.5$, this is not seen in the $BF,AD,AD$ and $D,AD,AD$ cases. 
For these last two cases, the greater $p_{\!_I}$ the lower the efficiency.  

We have also noted an important fact concerning the numerical optimization of the efficiency 
for the probabilistic protocols whenever the two qubits of the quantum channel are acted on by the amplitude
damping noise. In this situation the trade-off between efficiency and rate of success plays 
a crucial role in defining the range of values that $\theta$ (initial entanglement of the channel) 
and $\varphi$ (measuring basis) can assume during the numerical search for the optimal efficiency.  Indeed, if we allow
$\theta$ and $\varphi$ to run over all their possible values, i.e., from zero (no entanglement) to $\pi/4$ 
(maximal entanglement) and to $\pi/2$ (no entanglement), we are faced with solutions that give
very high values for the efficiency ($\overline{F}^j \approx 1$) while the rate of success 
is zero to the precision adopted in the maximization algorithm (8 numerical figures). The optimal $\theta$
in this case is almost zero, which means a quantum channel with almost no entanglement.
In order to avoid those unphysical solutions, we have restricted the ranges of $\theta$ and $\varphi$ 
to be such that $\theta_{min} \leq \theta,\varphi \leq \theta_{max}$. 
We observed that the more we restricted the range of $\theta$ and $\varphi$, 
the greater the probability of success and the lower the efficiency; and
when we set $\theta_{min}=\theta_{max}=\pi/4$, the probabilistic protocol gives
the same efficiency of the deterministic protocol.
The results presented in Fig. \ref{fig2}, in the circle-black curve of Fig. \ref{fig4}, and in Fig. \ref{fig5}
were obtained by setting $\theta_{min}=0.05\pi/2 = 0.07854$ and $\theta_{max}=0.95\pi/2=2.984$.
For all the other optimal results reported here, we have assumed  
$0 \leq \theta, \varphi \leq \pi/2$.

\begin{figure}[!ht]
\includegraphics[width=8.5cm]{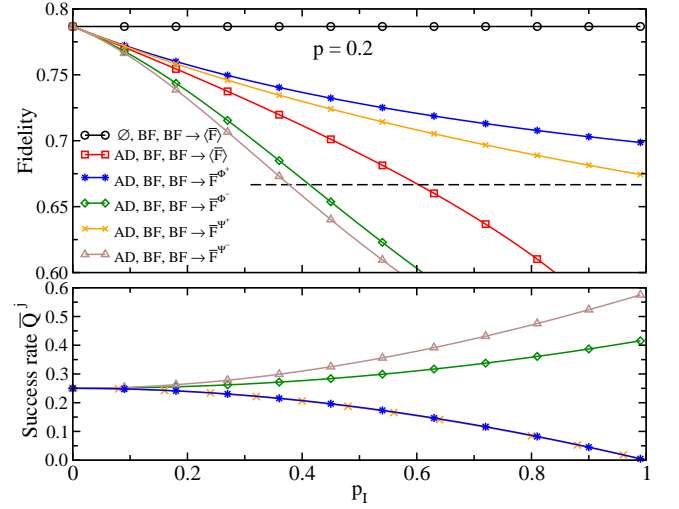}
\caption{\label{fig3}(color online) Top panel: Optimal efficiencies (average fidelities) for the deterministic and 
probabilistic protocols as a function of $p_{\!_I}$ for the noise arrangement involving the action
of the bit flip noise on the quantum channel and the amplitude damping noise on the input qubit.
Bottom panel: The probability of success associated to each one of the four possible measurement results
of Alice. The values of $\theta$ and $\varphi$ used to plot all $\overline{F}^j$ and $\overline{Q}^j$ 
are those that maximize $\overline{F}^{\Phi^+}$. The dashed-black curve marks the classical limit $2/3$ for the fidelity.}
\end{figure}

In Fig. \ref{fig3} we show the results obtained 
when we have the bit flip noise acting on the qubits
of the quantum channel and the amplitude damping noise acting
on the input qubit. The values of $\theta$ and $\varphi$ employed to draw
those curves are the ones optimizing $\overline{F}^{\Phi^+}_{AD,BF,BF}$. 
Now, contrary to the case where the amplitude
damping noise acted on the qubits of the channel, 
the optimal efficiency of the probabilistic protocol does
not surpass the optimal efficiency of the deterministic protocol when
no noise acts on the input qubit, i.e.,  $\overline{F}^j_{AD,BF,BF} < \langle\overline{F}\rangle_{\varnothing,BF,BF}$.
However, we still get that 
$\overline{F}^j_{AD,BF,BF} > \langle\overline{F}\rangle_{AD,BF,BF}$, showing that
the probabilistic protocol enhances the efficiency of the deterministic protocol under the same noise conditions.
Furthermore, for high values of $p_{\!_I}$ only the probabilistic protocol yields fidelities greater than
$2/3$, highlighting the importance of the probabilistic protocol in order to get a truly quantum teleportation.

\begin{figure}[!ht]
\includegraphics[width=8.5cm]{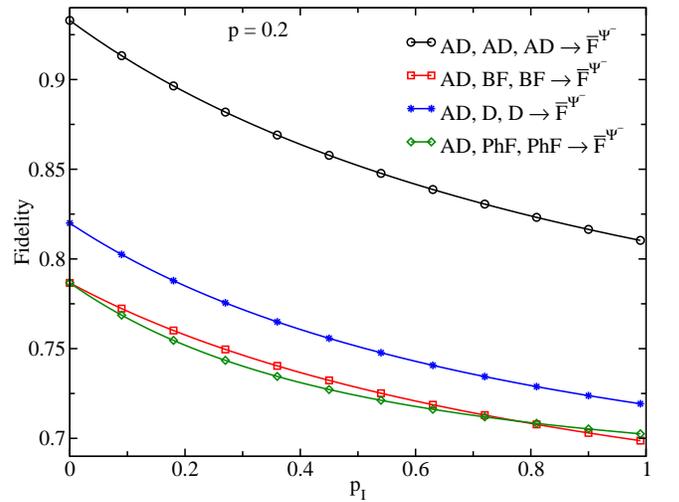}
\caption{\label{fig4}(color online) Optimal efficiencies for the probabilistic protocol when different types
of noise acts on the qubits of the quantum channel.}
\end{figure}

In Fig. \ref{fig4} we compare the optimal efficiency for a fixed type of noise acting on
the input qubit among all possibilities of noise acting on the quantum channel. The greatest 
efficiency occurs when all qubits suffer the amplitude damping noise. However, the probability of success
in this case is the lowest one. 

\begin{figure}[!ht]
\includegraphics[width=8.5cm]{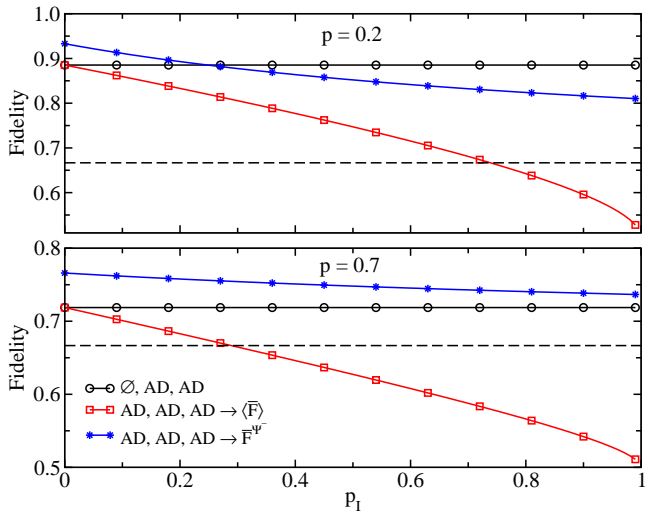}
\caption{\label{fig5}(color online)  Optimal efficiencies (average fidelities) as a function of $p_{\!_I}$ for the deterministic protocol
when only the quantum channel is subjected to the amplitude damping noise (circle-black curves) 
and for the deterministic (square-red curves) and probabilistic (star-blue curves)
protocols when all three qubits are subjected to the amplitude damping noise. 
The dashed-black curves mark the classical limit $2/3$ for the fidelity. The optimal efficiencies for the probabilistic protocol
are those obtained by post\-selecting $|\Psi^-\rangle$. For the two panels above, 
the success rate $\overline{Q}^{\Psi^-}$ is never lower than 
$0.4\%$ in the whole range of $p_{\!_I}$.}
\end{figure}

A very interesting noise arrangement is the one shown in Fig. \ref{fig5}, in which
all qubits are acted on by the amplitude damping noise (we continue to assume $p=p_{\!_A}=p_{\!_B}$).
The first thing worth noticing is that we can always find a $j$ such that $\overline{F}^j_{AD,AD,AD} > \langle\overline{F}\rangle_{AD,AD,AD}$
for any value of $p_{\!_I}$ and $p$, i.e, the probabilistic protocol always outperforms the
deterministic protocol under the same noise arrangements. We have also noticed that the optimal
$\overline{F}^{\Psi^{\pm}}_{AD,AD,AD}$ is always greater than the optimal 
$\overline{F}^{\Phi^{\pm}}_{AD,AD,AD}$. Another feature of this noise
arrangement is related to the fact that $\overline{F}^j_{AD,AD,AD} > \langle\overline{F}\rangle_{\varnothing,AD,AD}$
for the whole range of $p_{\!_I}$ whenever $p \gtrsim 0.5$ (lower panel of Fig. \ref{fig5}). 
For values of   
$p \lesssim 0.5$ we can also have the probabilistic protocol beating the deterministic protocol with no noise
acting on the input qubit. This only happens, however, when $p_{\!_I}$ is small (upper panel of Fig. \ref{fig5}).
Last, for certain values of $p$ and $p_{\!_I}$, the optimal parameters $\theta$ and $\varphi$ for a given
postselected measurement result also yield high average fidelities for other two possible measurement outcomes,
high enough to beat the optimal one of the deterministic protocol. In such cases Alice and Bob
can considerably increase the rate of success of the probabilistic protocol, and 
still outperform the efficiency of the deterministic protocol, by postselecting $3$ out of $4$
measurement results.

\subsection{Scenario $3$}

In this scenario the two qubits with Alice are acted on by the same type of noise during the same 
amount of time ($p_{\!_I}=p_{\!_A}=p$) and Bob's qubit is subjected to the same or a different type
of noise (see Fig. \ref{fig0}-c). Again, we have $16$ possible noise arrangements with only $6$ out of these
$16$ possibilities yielding probabilistic protocols with greater optimal efficiencies than the ones for the deterministic
protocols. Those $6$ cases contain the amplitude damping noise acting on 
Bob's qubit or on Alice's qubits. It is worth noticing that in this scenario there exists one case where 
the amplitude damping noise acts on Bob's qubit without yielding a better performance for the 
probabilistic protocol. In this case, where Alice's qubits are acted on by the phase flip noise ($PhF,PhF,AD$), 
both the probabilistic and deterministic optimal efficiencies coincide.  For the other $6$ cases in which
the amplitude damping noise is present the probabilistic protocol outperforms the deterministic one under
the same noise conditions. The qualitative behavior of these $6$ cases 
as well as their most important features are similar to the ones already
reported in scenario $2$. In particular, the optimal initial
entanglement of the quantum channel connecting Alice and Bob is not maximal.

It is important to mention that for all scenarios shown in Fig. \ref{fig0} and studied here we obtain nontrivial
probabilistic protocols, in the sense that they outperform the efficiency of the corresponding deterministic
protocols, if the amplitude damping noise is present. Whenever the amplitude damping noise is absent the 
efficiencies for the probabilistic and deterministic protocols coincide when optimizing the protocols as functions of
$\theta$ and $\varphi$, i.e., as functions of the initial entanglement of the quantum channel and of the type of projective
measurement implemented by Alice. In those cases where the coincidence occurs, the optimal $\theta$ is always the
one leading to the greatest initial entanglement ($\theta=\pi/4$). However, if we work with a one parameter optimization
problem ($\varphi$) and fix $\theta$ such that $\theta\neq \pi/4$, we can obtain probabilistic protocols 
outperforming the deterministic ones for noise arrangements in which the amplitude damping noise is not present.
In other words, if we are constrained from the start to work with non-maximally entangled quantum channels connecting Alice 
and Bob, other noise arrangements that do not include the amplitude damping noise lead to nontrivial probabilistic protocols.

\section{Conclusion}

We investigated the performance of the probabilistic (conditional) quantum
teleportation protocol  in the presence of noise. We have compared its optimal efficiency with the optimal one 
for the deterministic protocol under the same noise conditions.
We analyzed several noise arrangements in which the qubits employed in the execution of the teleportation protocol
are subjected to the most common types of noise encountered in the implementation 
of a quantum communication task, namely, the bit flip, the phase flip, the depolarizing,
and the amplitude damping noise.

For all noise arrangements here investigated, a total of $48$ distinct cases, only
$17$ cases have a probabilistic protocol with an optimal efficiency (average fidelity)   
greater than the optimal efficiency of the deterministic protocol.
We observed that a necessary condition for this to happen is that 
at least one of the qubits employed in the teleportation protocol 
must be subjected to the amplitude damping noise. Moreover, and similarly to the deterministic case,
for those $17$ noise arrangements the initial entanglement (prior to the action of noise) of the quantum channel connecting Alice and Bob leading to
the greatest efficiency is not maximal. This is
an example where \textit{less} entanglement means $\textit{more}$ efficiency, a feature already 
seen for deterministic protocols \cite{ban12,for15}.

We also showed several noise arrangements where \textit{more} noise
means \textit{more} efficiency. This happens whenever the efficiency of 
the probabilistic protocol, in which a certain number of qubits are subjected to noise, is
greater than the efficiency of the corresponding deterministic protocol with 
a noise arrangement where fewer qubits are acted on by noise. In addition to this we also
found situations in which \textit{different} noises mean \textit{more} efficiency. Indeed,
under certain noise arrangements we showed that it is better to have the qubits subjected to different types
of noise instead of the same noise in order to obtain the greatest efficiency.

We observed another important feature when comparing the optimal average fidelities of the probabilistic
and deterministic protocols under the same noise arrangement. In this scenario we found noise arrangements where
only the probabilistic protocol surpasses the classical threshold of $2/3$ for the average fidelity. This threshold 
means that a teleportation protocol yielding fidelities lower than $2/3$ can be simulated using only local operations
and classical communication (LOCC). Teleportation protocols with fidelities lying below this limit are not
considered truly quantum \cite{bra00}. Therefore, for some noise
arrangements we must employ the probabilistic instead of the deterministic protocol in order to obtain a 
quantum teleportation that is genuinely quantum.

Finally, for all the protocols here investigated we noted a trade-off between the 
rate of success and the efficiency. Indeed, the optimal protocols here reported
were obtained maximizing the average fidelity without any constraint on the value of
the probability of success. We can increase the rate of success, however, if we decrease 
the efficiency of the protocol. This is achieved 
by imposing a constraint on the lowest acceptable value for the probability of success.

\begin{acknowledgments}
RF thanks CAPES (Brazilian Agency for the Improvement of Personnel of Higher Education)
for funding and GR thanks the Brazilian agencies CNPq
(National Council for Scientific and Technological Development) and
CNPq/FAPESP (State of S\~ao Paulo Research Foundation) for financial support through the National Institute of
Science and Technology for Quantum Information.
\end{acknowledgments}

\end{document}